\documentclass[12pt]{article}
\def\be{\begin{equation}}
\def\ee{\end{equation}}
\def\la{\label}
\def\bea{\begin{eqnarray}}
\def\eea{\end{eqnarray}}
\def\non{\nonumber}
\def\ci{\cite}
\def\la{\label}
\def\bib{\bibitem}

\def\Lm{\Lambda}

\def\S{\Sigma}
\def\gm{\gamma}

\def\Omp{\Omega_\phi}
\def\Ompi{\Omega_{\phi i}}

\def\Om{\Omega}
\def\wp{w_\phi}

\def\Ompo{\Omega_{\phi o}}
\def\wpo{w_{\phi o}}
\def\weff{w_{eff}}
\def\s8{\sigma_8}

\def\fr{\frac}
\def\pp{\partial}

\usepackage{graphicx}

\begin{document}

\begin{flushright}
astro-ph/0203094 \\

\end{flushright}

\vspace{5mm}

\begin{center}
   {\Large \bf    Acceptable Inverse Power Law Potential Quintessence with $n=18/7$
   }

\end{center}

\vspace*{0.5cm}

\begin{center}
{\bf A. de la Macorra$^a$\footnote{e-mail: macorra@fisica.unam.mx} and G. German$^b$\footnote{e-mail: gabriel@fis.unam.mx}
}
\end{center}

\vspace*{0.1cm}

\begin{center}
{\small
\begin{tabular}{c}
$^a$ Instituto de F\'{\i}sica, UNAM\\ 
Apdo. Postal 20-364, 01000  M\'exico D.F., M\'exico\\
$^b$ Centro de Ciencias F\'{\i}sicas,
Universidad Nacional Aut{\'o}noma de M{\'e}xico,\\
Apartado Postal 48-3, 62251 Cuernavaca, Morelos,  M{\'e}xico\\
\\
\end{tabular}
}
\end{center}

\vspace{0.5cm}

\begin{center}
{\bf ABSTRACT}
\end{center}
\small{ We present a particle physics quintessence model  which agrees well with existing cosmological data, including the position of the acoustic peaks. This
model has  an inverse power law potential (IPL)  with $n=18/7\sim 2.57$ and
 it gives $\weff=-0.75$, an acoustic scale $l_A=307$ and a density contrast $\s8=0.95$. 

Models with $n>1$ have been said to be disfavored by the analysis of the acoustic
peaks. However, the results are not correct. The main reason is that the
 tracker approximation has been used in deriving the IPL constrains and for
$n<5$ the scalar field has  not reached its tracker value by present day.

The model can be   derived from particle physics, using 
Affleck-Dine-Seiberg "ADS" superpotential, for a  non-abelian 
gauge group with $N_c=8, N_f=1$. The advantage of having $N_f=1$ 
is that there is only one degree of freedom below the 
condensation scale given by the condensate (quintessence)
 $\phi^2=<Q\tilde Q>$ field.  The condensation scale is at $1 GeV$ a 
 very interesting scale since it connects the quintessence "Q" with the 
 standard model "SM"  scale. The  similarity in energy scales between $Q$ 
 and $SM$ scale gives an "explanation" to  the  coincidence problem. The fact 
that only recently the universe is accelerating  
is a natural consequence of the Q scale and the  evolution of $\phi$.

}


\noindent \rule[.1in]{14.5cm}{.002in}

\thispagestyle{empty}

\setcounter{page}{0} \vfill\eject

 The Maxima and Boomerang \ci{CMBR} observations on the cosmic
 radiation microwave background ("CMBR") and
the superonovae project SN1a \ci{SN1a}
have lead to conclude that the
universe is flat and it is expanding with an accelerating
velocity.  These conclusions show that the universe is now
dominated by an energy density $\Om_{\phi o}=0.7 \pm 0.1$ 
(the subscript "o" refers to present day quantities)
with  negative pressure.  The SN1a data requires 
an equation of state  $\wpo < -2/3$ \ci{w}  while recent
analysis on the CMBR peaks constrains the models to have
$\weff=-0.82^{+.14}_{-.11}$ \ci{neww,nww2}, where $\weff$ is an average
equation of state. This energy
is generically called the cosmological constant. Structure
formation also favors a non-vanishing cosmological constant
consistent with SN1a and CMBR observations \ci{structure}.
An interesting parameterization of this energy density is in
 terms of a scalar
field with gravitationally interaction only called quintessence
\ci{tracker}. The evolution of scalar field has been widely
studied and some general approach con be found in 
\ci{generic,mio.scalar,Wet}. The evolution of the scalar field 
$\phi$ depends on the functional form of its potential $V(\phi)$ 
and a late time accelerating universe constrains the form of the 
potential \ci{mio.scalar}.
 
One of the simplest and most interesting quintessence potentials
are the inverse power law (IPL) \ci{1/q}. In some special
cases they can be derived from non-abelian gauge theories 
\ci{bine,chris1,chris2} and
we can also have consistent models with a gauge coupling unified 
with the standard model (SM) couplings \ci{chris1}. 
 
 From the  CMBR analysis it has been inferred  that
IPL with $n<1$ are disfavored \ci{neww},\ci{wette}. 
However, in most cases one assumes a constant $w$ given by 
 the tracker value $w_{tr}=-2/(2+n)$ and for IPL models with $n<5$
the tracker solution \ci{tracker} is not a good approximation to the numerical 
(exact)
solution since the field has not reached its tracker value by present time.
This fact implies that we {\it cannot} use the tracker solution (i.e. 
the constant $w_{tr}$)
for the evolution of $\phi$  in  determining the
acoustic peaks. It is no surprise that the values of the acoustic peaks 
differ greatly if we use the tracker solution approximation or we evolve the
 quintessence field from its initial conditions. Models with $1<n<2.5$
 (for $\Ompi\geq 0.25$) are therefore still phenomenologically viable.

Tracker fields have the advantage that the value of $\wpo$ does 
not depend on the initial condition. In fact one can have more 
then 100 orders of magnitude on the initial conditions $\Ompi$ 
\ci{stein} and the value of $\wpo$ will not change.  In our case 
this is no longer so since the scalar field has not reached its 
tracker value and $\wpo$ depends slightly on $\Ompi$, so there is 
a dependence on $\Ompi$ but there is no fine tuning required on 
the initial conditions (we do not need to adjust the initial 
condition more than one significant figure). Furthermore, 
in our model, derived from non-abelian gauge theories, we can
determine the initial conditions in terms of the number of degrees
of freedom of the system. So we do not need 100 orders of magnitude
independence of initial conditions since they are well motivated and
given within one order of magnitude at most. Of course, any model 
which requires a fine tuning on the initial conditions would not be
theoretically acceptable but this is  not our case. In both cases, 
tracker and our model, one 
still has the coincidence problem since the scale $\Lm_c$ has to 
be tuned so that $\Omp\simeq 0.7$  with $h_o\simeq 0.7$ today. 
  
 There are  two constrains on the  equation of state parameter $\wp$, one
coming from direct observations SN1a which sets un upper limit, 
$\wpo<-2/3$ \ci{w} and the other is indirect and comes from 
numerical analysis of  the CMBR data and gives a smaller value 
$\weff=-0.82^{+.14}_{-.11}$ \ci{neww}.  Notice that the CMBR data 
gives a more negative $\wp$ than the SN1a one but it is an 
average equation of state (from last scattering to present day) 
while the SN1a result gives an $\wp$ in recent times. In IPL with 
$n<5$, where the quintessence field has not reached its tracker 
value yet, one has always $\wpo$ larger than $\weff$ in good 
agreement with SN1a and CMBR data. For  IPL models  it was shown 
\ci{chris2} that $\wpo$  depends on $n$ and the initial condition $\Ompi$. If 
we want  $\wpo < -2/3$  assuming an $\Ompi \geq 0.25$ IPL models 
require  an $n$ to be less  2.74  
\ci{chris2} assuming no contribution from radiation at present 
time. If we include radiation with $\Om_{ro}=4.17\times 10^{-5}h_o^2$ 
then the value of $n$ will decrease slightly, e.g for $\Ompi=0.25$ 
we have $\wpo\leq -2/3$ for $n\leq 2.5$. Larger values of $\Ompi$ allow 
larger values of $n$, however we would not expect to have $\Ompi$ 
much larger since a "reasonable" amount of energy must go into 
the standard model of elementary particles "SM". These results set an upper value of $n$ but there is 
still room for models with $1<n<2.5$ and if we take $\Omp\geq 
0.3$ then the value of $n$ can be as large as $n\leq 2.66$.  

The CMBR constrain on $\wp$ can be studied from the position of 
the third acoustic peak. The position of the third CMBR peak has been found to 
be not very sensitive to the different cosmological parameters
and it is a good quantity to obtain the acoustic scale $l_A$ \ci{doran}. The acoustic
scale  $l_A$, which sets the scale of the peaks,   derived from the third acoustic
peak is 
\be\la{la}
l_A=316\pm 8
\ee
were we have taken $l_3=845^{+12}_{-25}$ \ci{CMBR} 
(see below for the definition of $l_A$). 

Here we will present a model with the  largest value of $n$   
that is still in agreement with the observational cosmological 
data \ci{neww} and that can be nicely derived from particle 
physics. Since, for  a non-abelian gauge group (see below) we 
have  $n=2+4N_f/(N_c-N_f)=2+4/(z-1)$, where $N_f$ is the number
 of flavors of the gauge group $SU(N_c)$ and $z=N_c/N_f$, we can see 
that $n$ decreases with increasing $z$. The requirement on $n$ to 
be smaller than $2.66 \;(2.5) $ for $\Ompi\geq 0.3 \;(0.25)$,  in 
order to give the  correct phenomenology  and the observed 
values  of the acoustic scale and present day $\wpo$, implies 
that $z> 9\; (7)$ or $z\, N_f<N_c $. As we will see later the number 
of condensates of the gauge group $SU(N_c)$ is given in terms of 
$N_f$. Therefore, the model with the least number of condensates 
has $N_f=1$ and the smallest gauge group would be $N_c=8$. 
Furthermore, in string compactification for the heterotic string, 
the gauge group has at most rank 8 (as $SU(8)$). The value of
 $n$ for $SU(8)$ with $N_f=1$ is $n=18/7=2.57$. This model 
 represents the limiting acceptable model.

The acoustic scale  gives $l_A=307$ and $\wpo=-0.68$, which gives 
a good prediction of the acoustic peaks within observational 
limits. This acoustic scale should be compared with the tracker 
solution $l_{A tr}=281$ and  $w_{tr}=-2/(2+n)=-0.44$ and the 
cosmological constant $l_{A Cte}=315$ (i.e. $w_{Cte}\equiv-1$) 
for the same  initial conditions. We see that the tracking 
solution is not a good approximation since $\wpo$ differs by more 
than $38\%$ and the acoustic scale $l_A$ by $9\% $ from the  
numerical solution of the scalar field, discrepancy large enough 
to rule out the model.

Since our model has $n<5$  the quintessence field has not reached its
 tracker value yet and the coincidence
problem is not solved. However, there is a clear connection 
between  the model condensation scale $\Lm_c= 1 GeV$ and the 
standard model scale. So, we could think of the "solution" to the 
coincidence problem as the following: The scale of quintessence 
should not be given by today's energy density but by the 
condensation scale $\Lm_c$. The natural value of this scale is 
that of the standard model.  The subsequent evolution of the 
quintessence field is determined by the solution of the 
Friedmann-Robertson-Walker equations and the fact that only 
recently the universe is accelerating is a natural consequence of 
the quintessence dynamics starting at $\Lm_c$.

The model with $n=18/7$ can be easily obtained from  
   a $SU(8)$ non-abelian gauge  group with 
$N_f=1$ number of (chiral + antichiral) fields in the 
fundamental  representation and with a condensation scale 
$\Lm_c=1 GeV$, quite an interesting scale. Above the condensation 
scale the gauge coupling constant is small and the elementary 
fields are massless. At the condensation scale there is a phase 
transition, the gauge coupling constant becomes strong, and binds 
the elementary fields together forming meson fields. In this 
model there is only one  degree of freedom, $\phi$, in the 
confined phase and the Affeck-Dine-Seiberg "ADS" superpotential 
\ci{Affleck}  obtained is therefore exact.

At the beginning we have particles of the standard model (SM) and the quintessence
model (Q).  All fields, SM and $Q$ model, are
massless and redshift as radiation until we reach the condensation
scale $\Lm_c$ of Q group. Below this scale the fields of the quintessence
gauge group
will dynamically condense and we use ADS potential   to study
its cosmological evolution.  The ADS potential
is  non-perturbative and  exact (it receives no quantum corrections)
\ci{duality} and it is given 
for a non-abelian
$SU(N_c)$ gauge group with $N_f$ (chiral + antichiral) massless
matter fields  by  \ci{Affleck}
\be \la{super}
W=(N_c-N_f)(\fr{\Lm_c^{b_o}}{det <Q\tilde Q>})^{1/(N_c-N_f)}
\ee
where $b_o=3N_c-N_f$ is the one-loop beta function coefficient. 
The
 scalar potential in global supersymmetry is
$V=|W_\phi|^2$, with $W_\phi=\pp W/\pp \phi$, giving \ci{bine}
\be
V=c^2\Lm_c^{4+n}\phi^{-n}
 \la{v} 
 \ee
where we have taken $det <Q\tilde Q>=\Pi_{j=1}^{N_f}
\phi^2_j,\;\;c=2N_f,\;\;n=2+4\fr{N_f}{N_c-N_f}$
and $\Lm_c$ is the condensation scale
of the gauge group $SU(N_c)$.  Our model has $N_f=1, N_c=8$ and $n=18/7$. There are no baryons since $N_f<N_c$ and there is only one degree of freedom below $\Lm_c$ which is the
condensate $\phi=<Q\tilde Q>$.

The cosmological evolution of $\phi$ with an arbitrary potential
$V(\phi)$ can be determined from a system of differential
equations describing a spatially flat Friedmann--Robertson--Walker
universe in the presence of a barotropic fluid energy density
$\rho_{\gm}$ that can be either radiation or matter, are
\bea\la{eqFRW}
\dot H &=& -\frac{1}{2}(\rho_\gamma+ p_\gamma+\dot \phi^2),\non \\
\dot \rho &=& -3H(\rho+p),\\
\ddot \phi &=& -3H \dot \phi-\frac{d V(\phi)}{d \phi},\non
 \eea
where $H$ is the Hubble parameter ($H=100h\; km\,Mpc^{-1} 
s^{-1}$), $\dot f = df/d t$, $\rho$ ($p$) is the total energy 
density (pressure) and we are setting the reduced Planck mass 
$m^2_p=1/8\pi G\equiv 1$.

Solving eqs.(\ref{eqFRW}) we have that the
energy density of the $Q$ group $\Omp$ drops quickly,
independently of its initial conditions, and it is close to zero
for a long period of time, which  includes  nucleosynthesis (NS)
if $\Lm_c$ is larger than the NS energy $\Lm_{NS}$ (or temperature
$T_{NS}=0.1-10 MeV$), and becomes relevant only
until very recently \ci{chris2}.  On the other hand, if $\Lm_c <  \Lm_{NS}$ then
the NS bounds on relativistic degrees of freedom must be imposed
on the models. Finally, the energy density of $Q$ grows and it
dominates at present time the total energy density with
the  $\Ompo \simeq 0.7$ and a negative pressure $\wpo <-2/3$
leading to an accelerating universe \ci{w}.

The value of the condensation scale in terms of $H_o$ is \ci{bine,chris2}
\be \la{Lm}
\Lm_c=\left(\fr{3 y_o^2\phi^n H_o^2}{4N_f^2} \right)^\frac{1}{4+n}
  \ee
 The approximated value  can be obtained since
 one expects, in general, to have
 $  y_o^2 \phi_o^n \sim 1$
for a model with $\Ompo=0.7$ and $\wpo<-2/3$. The magnitude order of the
condensation scale is therefore $\Lm_c=H_o^{2/(4+n)}$.

In our model, the cosmological evolution requires a condensation
scale $\Lm_c=1 GeV$ in order to give  
$\Ompo=0.7\pm 0.1$ with a Hubble parameter $h_o=.65\pm.7$ at present time.
Since $\Lm_c \gg \Lm_{NS}$ the energy density at NS is $\Omp(NS) \ll 1$ 
and there is no constrain from nucleosynthesis on the model. 

In order to set the initial conditions for $\phi$ we will assume that all relativistic
degree of freedom (MSSM and Q) had the same fraction of energy density at high
energies, when all fields were massless.  The initial conditions could
be set at the unification scale or at the reheating temperature $T_{rh}$. If
$T_{rh}$ is larger then the supersymmetric masses (i.e. $T_{rh}> 10^3 GeV$),
which is a natural assumption, then all degrees of freedom (MSSM and Q)
would be relativistic at $T_{rh}$ and each degree of freedom would have the same energy density (assuming the standard reheating process which is gauge blind).
Therefore, the initial energy density conditions would be exactly the same 
at the unification scale or reheating temperature.

The MSSM has $g_{sm i}=228.75$  while the $Q$ group has $g_{Q i}= (1+7/8)(2(N_c^2-1)+2N_fN_c)=266.25$
degrees of freedom, where $g_{a}=\S_a Bosons+7/8 \S_a Fermions$. Taking into
account that some fields become massive at   lower energies, we can determine
the energy density at an arbitrary  energy scale $\Lm $ and it is given by \ci{chris1}
 \be\la{omq}
\Om_Q(\Lm) =\fr{g_{Q f}(g_{sm f}g_{Q i}/g_{sm  i}g_{Q f})^{4/3}}
{g_{sm  f}+g_{Q f}(g_{sm f}g_{Q i}/g_{sm  i}g_{Q f})^{4/3}} 
\ee
where $g_{sm i},\; g_{sm f}, \;g_{Q i}, \;g_{Q f}$ are the 
initial (i.e. at high energy scale) and final (i.e. at $\Lm$)   
standard model and $Q$ model relativistic degrees of freedom, 
respectively. Taking $\Lm=\Lm_c=1 GeV$ the MSSM has $g_{sm 
f}=10.75$ and if the $Q$ group is still supersymmetric at $\Lm_c$ 
it has $g_{Q f}=g_{Q i}$ and $\Ompi(\Lm_c)=0.29$. If $Q$ is no 
longer supersymmetric $g_{Q f}=g_{Q i}/2$ and 
$\Ompi(\Lm_c)=0.34$. So a reasonable choice for the initial 
conditions is $\Ompi(\Lm_c)=0.3$.  

We show in fig.(\ref{omw}) the evolution of $\Omp$ and $\wp$ as a 
function of $N=Log(a)$, with $a$ the scale factor, for initial conditions
$\Ompi=0.5,0.3,0.2$ short-dashed, long-dashed and 
solid lines, respectively. We see that $\wpo$
 decreases for larger initial condition $\Ompi$. For 
$\Ompi=0.9,0.5,0.3,.2$ one finds $\wpo=-0.95,-0.8,-0.68,-0.61$ 
and an acoustic scale
$l_A=314,313,307,303$, respectively. It is no surprise that for 
$\Ompi=0.2$ the value of $\wpo=-0.61$ lies outside the observed 
range $\wpo<-2/3$. This is because the model we are working with 
(i.e. n=18/7=2.57)  gives almost the limiting value of 
$\wpo=-2/3$  for $\Ompi= 0.3$ ( $\wpo$ increases for smaller $\Ompi$) and that was the reason for  using this model. 
However,  as long as we take $\Ompi\geq 0.27$ the model   
satisfies the cosmological constrains. For scalar fields that 
have reached its tracker value by present day (i.e. $n>5$) the 
initial value of $\Ompi$ is not constrained \ci{tracker} since it 
can vary for more than 100 orders of magnitude and the value of 
$\wpo$ will be the same $\wpo=-2/(2+n)$ \ci{tracker}, however
for $n>5$ one has $\wpo\geq -0.28$ which is too large.

\begin{figure}[p!]
\begin{center}
\includegraphics[width=9cm]{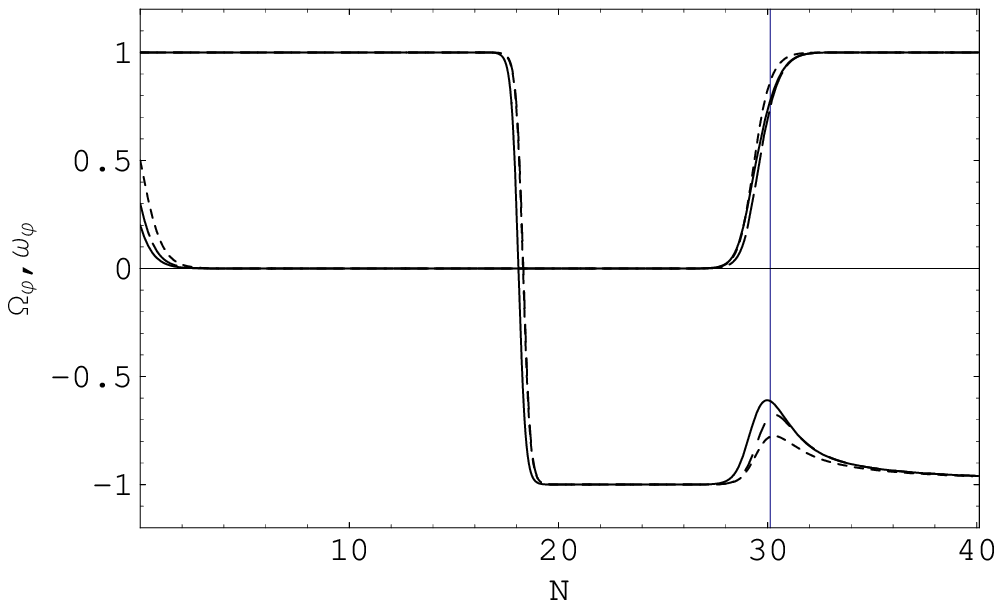}
\end{center}
\caption{\small{Variations on $\Ompi$ lead to different physical 
situations given by $\wpo$ and $\Ompo$. We have taken 
$\Ompi=0.5,0.3,.2$ short-dashed, long-dashed and solid lines, 
respectively.  The vertical line marks the time at $\Ompo=0.7$ 
with $h_0=0.67$. Notice that $\wpo$ increases with decreasing 
$\Ompi$. }}\la{omw}
\end{figure}

In previous works \ci{chris1,chris2} we have studied quintessence 
models that are unified with the standard model gauge groups, 
i.e. the gauge coupling constant of all gauge groups is the same 
at the unification scale. In the model we are working here, 
$N_f=1,\; N_c=8$,  the renormalization group equation given by 
$\Lm_{RG}=\Lm_{Gut}exp[-16\pi^2/2b_og_{gut}^2]=9\times 10^{12} 
GeV$ with $\Lm_{gut}=10^{16} GeV,\; g^2_{gut}=4\pi/25.7$ the 
unification scale and coupling \ci{unif}, respectively, and 
$b_o=3N_c-N_f=23$ the one-loop beta function coefficient. It is 
clear that $\Lm_{RG} \neq \Lm_c$ and so the model cannot be 
unified with the SM groups. If we insist in gauge coupling 
unification we would need, on top of the original 1 chiral + 1 
antichiral fields,  37 extra chiral fields. If all extra fields 
are chiral then they would not contribute to the ADS 
superpotential in eq.(\ref{super}).  Of course, we think that 
such a model is not natural but it is possible (4-D string models 
can have different number of chiral and antichiral fields).

The acoustic scale, that sets the scale of acoustic peaks, for 
a flat universe  is given by \ci{lA}
\be\la{la1}
l_A = \pi \fr{\tau_o-\tau_{ls}}{\bar c_s \tau_{ls}}
\ee
where $\tau_o$ and $\tau_{ls}$
are the conformal time today and at last 
scattering  ($\tau=\int dt a^{-1}(t), a(t) $ the scale factor) and $\bar c_s\equiv \tau_{ls}^{-1} \int_0^{\tau_{ls}} d\tau c_s$ is the average
sound speed before last scattering ($c_s^{-2}=3+(9/4)w_b(t)/w_r(t)$ with
$w_b=\Om_b h^2, \;w_r=\Om_r h^2$ the fraction of baryon and radiation energy
density, respectively).  The acoustic m-th peak $l_m$ is then given in 
terms of $l_A$ and a peak and model dependent  phase shift 
 $\varphi_m$, $l_m=l_A(m- \varphi_m)$. It has been observed
  in \ci{doran} that the third peak is quite insensitive to 
  different cosmological parameters that enter in determining
$\varphi_3\simeq 0.341$ and so the position of the third peak is
 a good quantity to extract
the acoustic scale $l_A$.  The data from Boomerang and Maxima
set the first three acoustic peaks, (the first through $l_{3/2}$)
 and the acoustic scale at \ci{CMBR}
\be\la{lp}
l_1=213^{+10}_{-13},\hspace{.5cm} l_2=541^{+20}_{-32},\hspace{.5cm}  
l_3=845^{+12}_{-25},\hspace{.5cm}  l_{3/2}=416^{+22}_{-32}, 
 \hspace{.5cm} l_A=316^{+8}_{-8}.
 \ee
We have solved eqs.(\ref{eqFRW}) numerically   with initial 
conditions $\Ompi(\Lm_c)=0.3$ at $\Lm_c=1 GeV$ imposing $h_o=0.65$, 
$\Ompo=0.75$ and we obtain  
\be\la{res1}
l_A=307, \hspace{.5cm}  \wpo=-0.68,\hspace{.5cm}  \weff=-0.75, 
\ee
where $\weff \equiv\int da\ \Om(a)\wp(a)/\int da\ \Om (a)$. 
The energy density al last scattering (LS) is negligible ($\Omp(LS)=10^{-9}$) 
and the average sound speed at LS  is $\bar c_s=0.52$. 
The result is not highly sensitive to the initial conditions and
a change in $\Ompi$ of $50\%$ will still be ok \ci{chris2}.

We see from eq.(\ref{res1}) that the acoustic scale $l_A$ is within the observational  range given by eq.(\ref{la}). The prediction of the first
three acoustic peaks and first through ($l_{3/2}$) is 
\be\la{res2}
l_1= 223,\hspace{.5cm}  l_2=542, \hspace{.5cm} l_3=829, \hspace{.5cm} l_{3/2}=414
\ee
for a baryon density $w_b=\Om_b h_o^2=0.02$ and $n_s=1$ the index
 of power spectrum of primordial density fluctuations.
The peak values in eq.(\ref{res2}) are consistent with 
the observational data in eq.(\ref{lp}) and we have used the phase shifts given in \ci{doran}.

The  value of $l_A$ is sensitive to $\Ompo$ and even more to $h_o$.
For increasing $h_o$ (with
$\Ompo$ fixed)
we find a decreasing $l_A$, e.g. for $h_o=0.6,\,0.65,\,0.7,\,0.8$ 
one has
$l_A=309,\,307,\,288,\,272$ respectively, while for an increasing 
energy density  one has an increasing $l_A$, e.g. $\Ompo= 0.6,\,0.7,\,0.75,\,0.8$ one gets $l_A= 291,\,300,\,307,\,317$ with fixed $h_o=0.65$. Since $\bar c_s$ depends on $w_b$ and an increase in $w_b$ makes
$c_s$ smaller and we have  therefore a slight increase in the 
acoustic scale, e.g.  for $w_b=.019,\,0.020,\,0.022,\,0.026$   one gets a 
value of $l_A=305,\,307, \,309,\,314$ with  $\Ompo=0.75,\, h_o=0.65$ fixed. 
A change in $n_s$ does not affect $l_A$ but
it changes the acoustic peaks through the phase shifts $\varphi_m$ slightly.

Another relevant cosmological quantity is the density contrast  
on scales of $8h^{-1} Mpc$, $\sigma_8$, which is constraint by 
the galaxies cluster abundance. For a flat universe the empirical 
fit of different authors (which converge within one $\sigma$) 
are: Eke et al. have $\s8=(0.52\pm 
0.08)\Om_m^{-0.52+0.13\Om_m}$\ci{eke}, Viana et al.  report their 
best fit  at $\s8=0.56\Om_{m}^{-0.47}(1 \pm 0.3)$ \ci{viana}, 
while Steinhardt et al. \ci{stein} have 
$\s8=[(0.5-0.1\,\Theta)\pm0.1]\Om_m^{-\gamma},\; 
\Theta=(n-1)+(h_o-0.65), \gamma=0.21-0.22w+0.25\Theta$. For $n=1, 
h_0=0.65$ one has $\s8=1.02 \pm .15,\,1.07 \pm 0.03,\,0.98 \pm 
0.2$ for Eke, Viana and Steinhardt respectively. Recent analysis  
give slightly lower values of   
$\sigma_8=(0.46^{+0.05}_{-0.07})\Omega_m^{-0.52}$ for \ci {hoek} 
(see also \ci{sigma8}) and depends quite strongly on $\Om_m$ (for 
smaller $\Om_m$ one has a larger $\sigma_8$). The central value 
for $\Om_m=0.25$ is $\sigma_8=0.945$ which agrees quite well with 
our the value obtained in our model $\s8=0.95$.  

To conclude, we have shown that inverse power law potentials with  $n>1$ are {\it not} disfavored with the existing cosmological data and we have shown an explicit
example with $n\simeq 2.57$.   In particular, the values of $\wpo$, the acoustic scale and  peaks and the density contrast $\s8$ lie within the observed data. The model has been derived from particle physics, using ADS superpotential,
from a non-abelian gauge group with  $N_c=8,\,N_f=1$.
Since $N_f=1$ there is only one degree of freedom below the condensation scale,
i.e. the condensate or quintessence field $\phi$. The condensation scale
is at $1 GeV$ a very interesting scale which connects quintessence with the standard model.

This work was supported in part by CONACYT project 32415-E and
DGAPA, UNAM project IN-110200.

\thebibliography{}

\footnotesize{

\bib{CMBR} {P. de Bernardis {\it et al}. astro-ph/0105292,
P. de Bernardis {\it et al}. Nature, (London) 404, (2000)
955, S. Hannany {\it et al}.,Astrophys.J.545 (2000) L1-L4}

\bib{SN1a} {A.G. Riess {\it et al.}, Astron. J. 116 (1998) 1009; S.
Perlmutter {\it et al}, ApJ 517 (1999) 565; P.M. Garnavich {\it et
al}, Ap.J 509 (1998) 74.}

\bib{structure}{ G. Efstathiou, S. Maddox and W. Sutherland,
 Nature 348 (1990) 705.
  J. Primack and A. Klypin, Nucl. Phys. Proc. Suppl. 51 B, (1996),
30}

\bib{w}{S. Perlmutter, M. Turner and  M. J. White,
Phys.Rev.Lett.83:670-673, 1999; T. Saini, S. Raychaudhury, V. Sahni
and  A.A. Starobinsky, Phys.Rev.Lett.85:1162-1165,2000 }

\bib{neww} C. Baccigalupi, A.  Balbi, S. Matarrese,
F. Perrotta, N.  Vittorio, astro-ph/0109097

\bib{nww2} R. Bean and A. Melchiorri, Phys. Rev. D 65, 041302(R) (2002), astro-ph/0110472.

\bib{wette}   M. Doran, M. Lilley, C. Wetterich  astro-ph/0105457

\bib{doran} M. Doran and  M. Lilley,   astro-ph/0104486 

\bib{lA}W.Hu, N. Sugiyama, J. Silk, Nature 386 (1997) 37
 
\bib{tracker} I. Zlatev, L. Wang and P.J. Steinhardt, Phys. Rev.
Lett.82 (1999) 8960;  Phys. Rev. D59 (1999)123504

\bib{1/q} {P.J.E. Peebles and B. Ratra, ApJ 325 (1988) L17;
Phys. Rev. D37 (1988) 3406}

\bib{bine}{P. Binetruy, Phys.Rev. D60 (1999) 063502, Int. J.Theor.
Phys.39 (2000) 1859,  A. Masiero, M. Pietroni and F. Rosati, Phys.
Rev. D61 (2000) 023509}

\bib{chris1}{A. de la Macorra and C. Stephan-Otto,  Phys.Rev.Lett.87 (2001) 271301,
A. de la Macorra hep-ph/0111292  }

\bib{generic}  A.R. Liddle and R.J. Scherrer, Phys.Rev.
D59,  (1999) 023509

\bib{mio.scalar}{A. de la Macorra and G. Piccinelli, Phys.
Rev.D61 (2000) 123503}
\bib{Wet}{C. Wetterich, Astron. Astrophys.301 (1995) 321

\bib{Affleck}{I. Affleck, M. Dine and N. Seiberg, Nucl. Phys.B256
(1985) 557}

\bib{chris2} A. de la Macorra and C. Stephan-Otto,  Phys.Rev.D65:083520,2002 

\bib{unif}{U. Amaldi, W. de Boer and H. Furstenau, Phys. Lett.B260
(1991) 447, P.Langacker and M. Luo, Phys. Rev.D44 (1991) 817}
}

\bib{duality}{K. Intriligator and  N. Seiberg, Nucl.Phys.Proc.Suppl.45BC:1-28,1996

\bib{eke} V.R. Eke, S.Cole,  and C.S. Frenk, MNRS, 282, 263 (1996) 

\bib{viana} P.T.P. Viana   and A.R. Liddle, astro-ph/9902245. To appear in the electronic proceedings of the conference
Cosmological Constraints from X-ray Clusters, Strasbourg, France, Dec. 9-11, 1998

\bib{stein} L. Wang and P. J. Steinhardt   Astrophys.J.508:483-490,1998
\bib{hoek} H. Hoekstra, H. Yee, M. Gladders  
astro-ph/0204295
\bib{sigma8}
P.  Schuecker, H. Bohringer, C. A. Collins, L. Guzzo.  
astro-ph/0208251;   G. Holder , Z. Haiman,  J. Mohr   
Astrophys.J.560:L111-L114,2001 (astro-ph/0105396)

\end{document}